\documentclass[prl,aps,cite]{revtex4}

\newcommand{\newc}{\newcommand}
\newc{\beq}    {\begin{equation}}
\newc{\eeq}    {\end{equation}}
\newc{\beqa}    {\begin{eqnarray}}
\newc{\eeqa}    {\end{eqnarray}}
\newc{\bs}    {\section}
\newc{\no}    {\\ \nonumber}

\def\PLA{{ Phys. Lett.} {\bf A} }

\def\PRL{{ Phys. Rev. Lett. }}

\def\PRA{{ Phys. Rev.} {\bf A} }

\usepackage{graphicx}

\begin{document}
\title{Quantum key distribution  using superposition of the vacuum and single photon states}
\author{Jae-Weon Lee and  Jaewan Kim }
\address{
School of Computational Sciences, Korea Institute for Advanced
Study, 207-43 Cheongryangri-dong, Dongdaemun-gu Seoul 130-012,
Korea}
\author{  Yong Wook Cheong }
\address{Quantum Photonic Science Research Center, Hanyang University, Seoul
133-791, Korea}
\author{  Hai-Woong Lee}
\address{Department of Physics, Korea Advanced Institute of Science and
Technology, Daejon 305-701, Korea }
\author{  Eok Kyun Lee  }
\address{Department of Chemistry,  School of Molecular Science (BK 21),
Korea Advanced   Institute of Science and Technology,  Daejon
 305-701, Korea.}

\date{\today}

\begin{abstract}
{\normalsize B92-type and BB84-type  quantum cryptography schemes
using superposed states of the vacuum and single particle states
which are robust against PNS attacks are studied. The number of
securely transferred classical bits per particle (not per qubit)
sent in these schemes is calculated and found to have upper
bounds.
 Possible experimental realizations using the cavity QED or linear optics are suggested.
}
\end{abstract}
\maketitle
%
\bs{I. Introduction}

 Recent progress in theories and
 experiments\cite{PRL89,PRA62,knill,experiment,leeyh}
of generation and manipulation of single photons
allows one to think the quantum information processing
utilizing single particles feasible.
 The first
commercial application of quantum information science at a single
qubit level might be the quantum key distribution
(QKD)\cite{gisin}. In the typical QKD scheme a sender (Alice)
shares a secret key with a receiver(Bob) by  sending superposition
of photon polarization states. However, one can encode information
not only in  particle states but also in the vacuum as shown in
some QKD schemes (mainly in double ray
schemes)\cite{B92,ardehali,qcomp}. In other words the vacuum can
play a role of an information carrier as particles do. Two of
authors had suggested the quantum teleportation and the Bell
inequality test using single-particle entanglement which was
verified experimentally later\cite{leekim}. In this direction we
proposed\cite{myepr} the Ekert-type \cite{Ekert} single ray
quantum cryptography scheme  using the entangled states of the
vacuum and the single particle state\cite
{czachor,bjork,Tan,hardy,santos,mann,gerry,michler,leekim,giorgi}.
The main purpose of this work is  to present a single ray B92-type
\cite{B92} and a BB84-type\cite{BB84} quantum cryptography schemes
using superposed states of  the vacuum and single photon states.
In our schemes the detection of the superposition state is
possible with cavity QED devices or linear optics devices  and
single photon detectors. Consider a quantum memory with the state
$|\phi\rangle=\alpha|0\rangle+\beta|1\rangle$ which is a
superposition of the vacuum $|0\rangle$ and a single photon state
$|1\rangle$ (optical qubit) which can be prepared by a photon
source using parametric down conversion\cite{hardy} or linear
optics\cite{lund} with the optical state truncation. By choosing
$|\beta|$ small enough we can make the expected number of photons
in $|\phi\rangle$ (i.e., $|\beta|^2$) arbitrary small. It implies
that we can encode classical bit information in the superposition
of  the vacuum and single photon state with $|\beta|\ll 1$, which
is  very faint light. Does this also mean that the legitimate
participants can share their secret key through a QKD protocol
with an arbitrary  faint light source to hide quantum channel
itself from eavesdroppers? We show that the answer is $no$ at
least  for straight forward generalizations of  B92 and BB84-type
QKD with the vacuum-photon superposed state considered in this
paper, and there are upper bounds for classical bits shared
between parties per particle  sent ($K$ defined below, not
counting the vacuum) in these schemes.

This paper is organized as follows.
In Sec. II we present a B92-type quantum cryptography scheme using the superposition of
the vacuum and single particle states. We also calculate the number of classical bits
transferred per particle.
In Sec. III we extend the arguments to a BB84-type scheme.
In Sec. IV possible experimental realizations of our schemes using cavity QED
and linear optics are presented, and a security analysis is given.
Finally, in Sec. V we present a concluding discussion.
\begin{figure}[htbp]
\includegraphics[width=7cm,height=5cm]{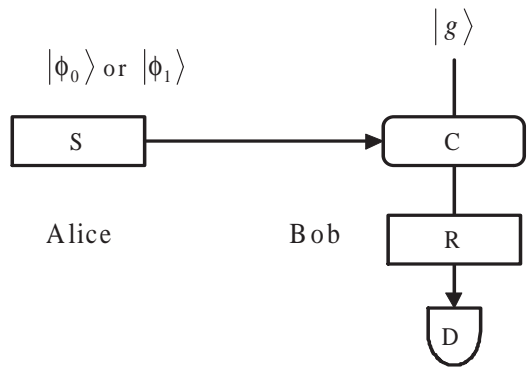}
\caption[Fig1]{\label{fig1}
Schematic of the cavity QED apparatus used in
the B92-type quantum cryptography scheme  using superpositions of the vacuum and the single photon state.
See text for detailed explanations.
}
\end{figure}

\bs{II. B92-Type Scheme}
Fig. 1 shows our
B92-type quantum cryptography scheme using a cavity QED device.
As is well known the B92 protocol exploits the fact that arbitrary two non-orthogonal states
can not be distinguished perfectly.
Basically our  scheme  with  a superposition of the vacuum and single photon
is just the same as the B92 scheme except for the state and the measuring device used.
For clarification we describe the scheme.

 (i) Alice sends sequences of states randomly chosen
between two non-orthogonal states $|\phi_0\rangle$ and
$|\phi_1\rangle$ representing logical 0 and 1, respectively; \beqa
|\phi_0\rangle&=&\alpha_0|0\rangle+\beta_0|1\rangle,\no
|\phi_1\rangle&=&\alpha_1|0\rangle+\beta_1|1\rangle, \eeqa
with normalization $|\alpha_i|^2+|\beta_i|^2=1~(i=0,1)$.\\
(ii) At a photon arrival time Bob measures a projection
operator randomly chosen between $P_0$ and $P_1$;
\beqa
\label{p}
P_0&\equiv &1-|\phi_1\rangle\langle\phi_1|,\no
P_1&\equiv &1-|\phi_0\rangle\langle\phi_0|.
\eeqa
(iii) After a series of measurements  Bob publicly
announces to Alice  in which instances he obtained a positive result.
This happens only when Alice sends $|\phi_0\rangle$
and Bob measures $P_0$ or Alice sends $|\phi_1\rangle$
and Bob measures $P_1$.
In other words, with probability  $1/2$ the state sent by Alice and
the projection operator are correlated.
In these cases,
applying projection $P_0~(P_1)$  to  $|\phi_0\rangle~(|\phi_1\rangle)$, Bob
 obtains a positive result
with a probability\cite{POVM}
\beq
p=\frac{1}{2}(1-|\langle \phi_0|\phi_1\rangle|^2)=
\frac{1}{2}(1-|\alpha_0^*  \alpha_1+\beta_0^*  \beta_1|^2).
\eeq
Thus, after $N$ trials, the total $n_b\equiv pN < N$ bits of keys
are successfully shared, if there have been no eavesdropping or errors.

(iv) To certify the absence of an eavesdropper Alice and Bob
sacrifice parts of data to check whether Bob obtained positive
results on $P_0~(P_1)$ measurement or not even in the case that
Alice sent $|\phi_1\rangle~(|\phi_0\rangle)$.

At this point one can pose an interesting question.
How many classical bits  can Alice
transfer per particle sent to Bob in an ideal case without errors or eavesdropping
up to the step (iii)?
 According to the Holevo's
theorem\cite{holevo,capacity} asymptotically
one cannot encode
and retrieve reliably more than one bit of
classical information per $qubit$.
Note, however,
 that in this letter we are interested in classical bit information  not per qubit but per particle excluding
the vacuum, so
the bit information per $particle$ is not restricted by the Holevo bound.
In our schemes the number of qubits sent  is not
equal to the number of particles sent, because the states sent are superpositions  of  the vacuum and
single particle states.
Let us calculate the ratio $K$.
Since Alice should choose  randomly  between
 $|\phi_0\rangle$ and $|\phi_1\rangle$,
the density matrix for a transmission
can be written as $\rho=(|\phi_0\rangle \langle \phi_0|+
|\phi_1\rangle \langle \phi_1|)/2$.
So the average number of particles sent to Bob
is
\beq
n_p= N~Tr( \rho{\hat{n)}}=\frac{N}{2}(|\beta_0|^2+|\beta_1|^2) \le N
\eeq
where  $\hat{n}$ is the particle number operator.
Then the ratio of bits transferred successfully to the average
number of  particles
sent is therefore
\beq
K\equiv\frac{n_b}{n_p}=\frac{1-|\alpha_0^*  \alpha_1+\beta_0^*  \beta_1|^2}{
|\beta_0|^2+|\beta_1|^2}.
\eeq
Without loss of generality, using
the Bloch representation $(\alpha_i,\beta_i)=(cos(\theta_i),e^{i\psi_i}sin(\theta_i)),~(i=0,1)$
we can rewrite $K$ as
\beqa
\label{K}
K&=&\frac{1-|cos(\theta_0)cos(\theta_1)+sin(\theta_0)sin(\theta_1)
e^{i\psi}|^2}{sin^2(\theta_0)+sin^2(\theta_1)}\no
&\le&
\frac{1-\left (|cos(\theta_0)cos(\theta_1)|-|sin(\theta_0)sin(\theta_1)|\right )^2}{sin^2(\theta_0)+sin^2(\theta_1)}
\no
&\equiv& K_{max} (\theta_0,\theta_1)
,
\eeqa
where $\psi\equiv\psi_1-\psi_0$ and the equality of the second line is satisfied when $\psi=0$ or $\pi$ and $
cos(\theta_0)cos(\theta_1)$ is opposite in sign to $sin(\theta_0)sin(\theta_1)e^{i\psi}$.
The upper bound on $K$ achieves value  2 asymptotically when  $sin(\theta_0)=\pm sin(\theta_1) \rightarrow 0$
(but still nonzero, See Fig. 2).
Note that the optimal states are near the vacuum but not the vacuum states.
In the case that  only one of $sin(\theta_0)$ and $sin(\theta_1)$ is $0$
(i.e., when one of $|\phi_0\rangle$ and $|\phi_1\rangle$ is the vacuum state.), $K$ becomes $1$.
$K=2$ means that in our QKD scheme Alice and Bob can share secure 2 classical bits of information
per single quantum particle transfer on the average in the ideal situation
of no  error or eavesdropping. The physical reason for this bound is that
the more we send the vacuum (i.e. $|\beta_i|\simeq 0$), the smaller $n_p$ is, but then it is harder to
 distinguish $|\phi_1\rangle$ from $|\phi_0\rangle$.
One can improve the probability of correct classification into
\beq p=1-|\alpha_0^*  \alpha_1+\beta_0^*  \beta_1| \eeq by optimal
positive operator valued measures (POVM)\cite{POVM,POVM2} using
ancilla qubits. In this case a similar argument leads to \beq
K=\frac{n_b}{n_p}=2\frac{1-|\alpha_0^*  \alpha_1+\beta_0^*
\beta_1|}{ |\beta_0|^2+|\beta_1|^2}, \eeq which  also has a
maximum value $2$ under the same condition stated below  Eq.
(\ref{K}). (However, we will not consider a specific  realization
of the POVM measurement
  in this paper.)
On the other hand,  for ordinary QKD
 schemes using ordinary particle states
 $K$ is usually smaller than $1$, because
there are always discarded data (i.e., $n_b < N$) to prevent Eve from distinguishing
Alice's states perfectly, while
 to represent a qubit one or more particles are required(i.e., $n_p\ge N$).
For example the typical  B92 scheme using two non-orthogonal photon polarization states has
$n_b=N/2$ (because the probability to get correct sifted keys is 1/2) and $n_p=N$, hence $K=1/2$.
In this sense, one can say that our B92-type scheme requires
a relatively smaller  (four times smaller) number of particles to be transferred to send
 a given classical bit information
 than the typical B92  quantum cryptography
schemes.
\begin{figure}[htbp]
\includegraphics[width=8cm,height=7cm]{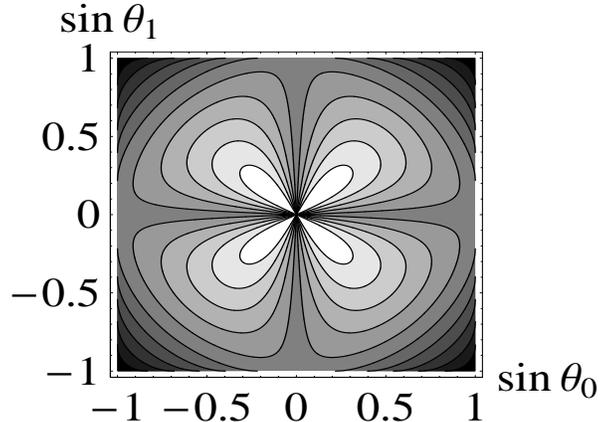}
 \caption[Fig2]{\label{fig2}
 Contour plot of the upper bound on the $K$ values ($K_{max}$) as a function of
 $sin(\theta_0)$ and $sin(\theta_1)$.
Gray level represents the
value from black(0) to white(2).
}
\end{figure}
This ratio $K$ is similar to the key ratio per energy \cite{energy,cabello} but not exactly equal to that,
because  in QKD schemes usually a sender and a receiver consume additional energy to share their references
and to communicate publicly.
$K$ is  more likely a ``key rate per brightness" which measure how dark the quantum channel is for
a fixed information transmission rate.
If we have a QKD scheme with very big $K$, we can use the scheme to hide the quantum
 channel itself from eavesdroppers, who will encounter a problem to find out where and when the
 quantum channel opens.
This fact can be especially useful for QKD schemes on a moving
satellite or free-space systems. The bound for $K$ for our scheme
is non-trivial. If $K=2$ is just the Holevo bound divided by $1/2$
(vaguely guessed proportion of particles in the vacuum-single
photon superposed states), then we should also have  maximal $K=2$
for the quantum memory or ordinary quantum channel but this is not
the case (maximal $K=\infty$ for these cases). Furthermore the
optimal $K$ value for our scheme even does not corresponds to the
case where
 the average particle number in the state is 1/2.
Meanwhile the fact that one can store or send multi-bits
information per particle is not so surprising. In fact, it is
shown that  infinite information can be transferred through
quantum channel at the cost of infinite entropy\cite{energy}.
However, what is interesting here is that although one can store
or send infinite information per particle in the quantum memory or
through a quantum channel using the vacuum-single particle
superposition, there is a upper bound for the key rate per
particle using QKD schemes
  with the superposition ( at least for the QKD schemes considered in this paper).

 \bs{III. BB84-Type Scheme}

Compared to the B92 scheme, the BB84 scheme is known to be
 more robust against the state discrimination attack\cite{usd}.
 It is straightforward to extend our consideration to the BB84-type scheme\cite{BB84}
 with  superposed  states of the vacuum and single particles.
 As in the typical BB84 scheme,
 Alice sends one of four states from two classes $\{|\phi_0\rangle,|\phi_1\rangle\}$, and $\{|\phi'_0\rangle$
 $|\phi'_1\rangle\}$ to Bob where
 $\langle \phi_0|\phi_1\rangle=0= \langle \phi'_0|\phi'_1\rangle$
 and $|\phi_i\rangle~(i=0,1)$ are not orthogonal to
$|\phi'_i\rangle$. Then, Bob measures one of four projection operators
$P_i=|\phi_i\rangle \langle \phi_i|$ or $P'_i=|\phi'_i\rangle \langle \phi'_i|$.
After basis reconciliation with Alice via a public channel Bob would get the classical bit
information with probability 1/2, hence $n_b=N/2$. The density matrix of Alice's particle
is
$\rho=\left(|\phi_0\rangle \langle \phi_0|+
|\phi_1\rangle \langle \phi_1| +|\phi'_0\rangle \langle \phi'_0|+
|\phi'_1\rangle \langle \phi'_1|\right)/4$.
Therefore, the ratio of bits shared to the number of  particles sent in this scheme is
\beq
K=\frac{4}{2(|\beta_0|^2+|\beta_1|^2+|\beta'_0|^2+|\beta_1'|^2)},
\eeq
where $|\phi_i\rangle=\alpha_i|0\rangle+\beta_i|1\rangle$
and $|\phi'_i\rangle=\alpha'_i|0\rangle+\beta'_i|1\rangle$.
The orthogonality condition  $\langle \phi_0|\phi_1\rangle=0=\langle \phi'_0|\phi'_1\rangle$ implies
$|\beta_0|^2+|\beta_1|^2=1=|\beta'_0|^2+|\beta'_1|^2$, so $K=1$ which is twice  the value of
$K$ for BB84 schemes with ordinary particles, because for the ordinary BB84 protocol
$n_b=N/2$ and $n_p=N$.

\bs{IV. Apparatus and security}

We may now proceed to the description of
 the apparatuses for our schemes shown in Fig. 1 for the B92-type scheme  and in Fig. 3 for the BB84-type
 scheme.
The setups consist of Alice's  photon source(S) for generation of
the superposition of the vacuum and  single particle states, and
Bob's projective  measurement device using either cavity QED (Fig.
1) or linear optics (Fig. 3) which are considered by many
authors\cite{davidovich,moussa,freyberger}. In principle the
cavity QED devices and the linear optics devices can be used both
for the B92 or the BB84 scheme. The detectors are essentially the
same detectors we considered in our previous work\cite{myepr}, so
we will just briefly review here. Let us first consider Fig. 1. By
utilizing the parametric down conversion or coherent light, the
source(S) generates $\phi_0$ or $\phi_1$ on Alice's demand.
Assuming  that at time $t=0$ a ground state atom
 $|g\rangle$
is injected into the cavity $C$,
the total cavity-atom state is then
$|\psi(0)\rangle=|\phi_i\rangle|g\rangle$.
The interaction between atoms and photons
in the cavity $C$ are described by
the Jaynes-Cummings Hamiltonian.
With this Hamiltonian and by choosing interaction time appropriately
one can  transfer the information  of photon
states $|\phi_i\rangle$
 to that of the atoms (See the references for details). Then the projective measurement on the photon
state $\alpha|0\rangle+\beta|1\rangle$
can be possible by adjusting  appropriately
the field in  the Ramsey zones ($R$)
such that the  state
undergoes a unitary evolution to the state
which  registers a click in  the state-selective
 ionization detector $D$\cite{davidovich,gerry}. So this setup
 performs ultimately deterministic (i.e., with probability 1 for ideal cases)
 projection on $|B\rangle\equiv\alpha|0\rangle+\beta|1\rangle$.
Let us find $|B\rangle$ such that   $P_0$ in Eq. (\ref{p})
can be written as
$|B\rangle \langle B|$.
It should satisfies $\langle B| \phi_1 \rangle =0$,
 because $P_0|\phi_1\rangle=0$.
 In other words, to measure $P_0$
 Bob should set the fields in the Ramsey zone
 so that the input photon state with
 $\alpha=\beta^*_1$ and $\beta=-\alpha^*_1$
  (i.e. orthogonal to $|\phi_1\rangle$)
 corresponds to  the click on detector $D$.
Similarly Bob can measure $P_1$ by performing
projection on $\beta^*_0 |0\rangle -\alpha^*_0|1\rangle$.
\\

\begin{figure}[htbp]
\includegraphics[width=8cm,height=5cm]{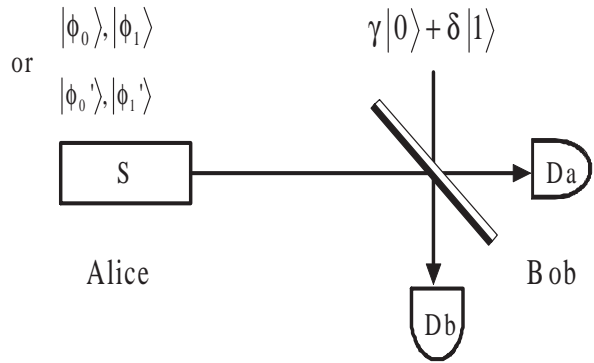}
 \caption[Fig3]{\label{fig3}
Schematic of the linear optics apparatus used in
the BB84-type quantum cryptography scheme  using the superpositions of the vacuum and
the single particle state.
See text for detailed explanations.
}
\end{figure}
On the other hand, the projective measurement for the BB84-type scheme using linear optics shown in Fig. \ref{fig3} is
non-deterministic in a sense that the measurement succeeds only  probabilistically.
 This setup is a modification of
 the setup  proposed in ref. \cite{lund}.
The beam splitter $BS$
performs the mode transformation
\beq
\label{trans}
  \left (
         \begin{array}{cc}
          a'\\
          b'
           \end{array}
    \right )=\left (
         \begin{array}{cc}
          \sqrt{R} & \sqrt{1-R}  \\
          -\sqrt{1-R} & \sqrt{R}
           \end{array}
    \right )
    \left (
         \begin{array}{cc}
          a\\
          b
           \end{array}
    \right ),
\eeq
where $R$ is the reflectivity of the beam splitter.
In  second quantized notation, the general input state
shown in  Fig. \ref{fig3} can be
written  as
\beq
\label{bs1}
\psi=(\gamma+ \delta~a^{\dagger})
(\alpha+\beta~b^{\dagger})|0\rangle
\eeq
with normalization  $|\alpha|^2+|\beta|^2=1$.
Here, $(\gamma+\delta~ a^{\dagger})|0\rangle$ is a known probe state, while
$\alpha|0\rangle+\beta|1\rangle$ is an unknown input state to be
measured.
By replacing $a$ and $b$ in
Eq. (\ref{bs1}) with $a'$ and $b'$ using Eq. (\ref{trans}),
 we obtain the output state
$
\psi=[
  \alpha\gamma  + \frac{\beta\delta}{2}
   (a'^{\dagger2}- b'^{\dagger2})
   +   \sqrt{2}\beta\gamma a'^{\dagger}
   ]|0\rangle,
   $
if we set   $R=1/2$ and $\alpha\delta= \beta \gamma$. Therefore,
by noting that the detector $D_a$ detects single photon and $D_b$
detects none, Bob can perform projective measurement on  the
superposition state $\alpha|0\rangle +\beta|1\rangle$ with the
probability of success $2|\beta \gamma|^2\le 1/2$\cite{myepr}. To
detect $|1\rangle$ or $|0\rangle$ state we simply replace the beam
splitter and check whether the detector $D_a$ fires or not. Then,
one of four states
$\{|\phi_0\rangle,|\phi_1\rangle,|\phi'_0\rangle$
 $|\phi'_1\rangle\}$ of the section III sent to Bob by Alice can be measured with this apparatus
 for the BB84-type scheme.

Let us now discuss the security of our schemes.
Basically our schemes follow the ordinary B92 and the BB84 schemes except for
the states and measuring devices used, so one can
 simply adopt the well known  security proof
for these ordinary schemes\cite{security} for our schemes also.
Another merit of our schemes is that since the superposition of the vacuum and one photon is
not a photon number eigenstate, our schemes are robust against the photon number
splitting (PNS) attacks\cite{pns}. (The PNS attacks  restrict key rates and distance  for many
 practical QKD schemes such as typical B92 or BB84 QKD schemes with weak coherent states.)
Because, even in the case  the eavesdropper (Eve) has multiple
copies of the state ($|\phi\rangle^{\otimes n}$) due to
imperfections of the light sources,
  to do the PNS attack Eve should perform the photon number non-demolition measurement,
but our schemes use the superposed states of the vacuum and single
photon which is inevitably destroyed by any photon number
measurement. This allows Alice and Bob to detect Eve attempting
the PNS attack by publicly comparing parts of the qubits sent with
the qubits received. So our schemes present yet another way for
security against the PNS attacks different from the recently
proposed schemes\cite{newschemes}.

For our schemes sending the pure vacuum ($|\phi\rangle=|0\rangle$) as a qubit has an intrinsic problem that
Bob can not distinguish the vacuum from channel loss.
But fortunately as described above  the pure vacuum state is not the optimal
state for the maximal $K$ value. So there is no  reason
  to use the pure vacuum state as a qubit for our schemes and we
can avoid this problem by simply not using the pure vacuum state.
In a practical sense, it is experimentally interesting but
challenging  to implement the detection of a superposition of  the
vacuum and single-photon states\cite{lund,singledetect}. Recently,
there are many related experimental and theoretical
 works about transferring  quantum states using the cavity\cite{cavityexp}.

\bs{V. Discussion}
In summary, we have proposed the  B92 and the BB84-type
quantum key distribution schemes using  superposed states of the vacuum and  the single particle state
robust against PNS attacks.
 We showed that in our QKD schemes using the vacuum-photon superposition states
  the information transferred  per
 particle sent is bounded. So far it is unclear that this restriction has more profound physical reasons.
 Therefore proving or disproving the existence of an exotic QKD protocol which
 has a big value of $K$, that is, a QKD scheme with  very faint light might
 be an interesting subject.
\\
\\
\indent J. Lee was supported by part by the Korea Ministry of Science and Technology.
J. Kim
was supported by the Korea Research Foundation (Grant
No. KRF-2002-070-C00029).
 H. W. Lee was supported by the Ministry of Science and Technology of Korea.
E. Lee
was supported by the Korea Research Foundation (
Grant No. KRF-GH16110).


\end{document}